\documentstyle[epsf,graphicx]{mn}

\def\etal{{et al.\ }}

\def\g{$\Gamma$}
\def\x2{$\chi^{2}$}

\def\asca{{\it ASCA }}
\def\rosat{{\it ROSAT }}

\def\einstein{{\it EINSTEIN }}

\def\x2{$\chi^{2}$}

\newbox\grsign \setbox\grsign=\hbox{$>$} \newdimen\grdimen \grdimen=\ht\grsign
\newbox\simlessbox \newbox\simgreatbox \newbox\simpropbox
\setbox\simgreatbox=\hbox{\raise.5ex\hbox{$>$}\llap
     {\lower.5ex\hbox{$\sim$}}}\ht1=\grdimen\dp1=0pt
\setbox\simlessbox=\hbox{\raise.5ex\hbox{$<$}\llap
     {\lower.5ex\hbox{$\sim$}}}\ht2=\grdimen\dp2=0pt
\setbox\simpropbox=\hbox{\raise.5ex\hbox{$\propto$}\llap
     {\lower.5ex\hbox{$\sim$}}}\ht2=\grdimen\dp2=0pt

\begin{document}

\title[X-ray luminous star-forming galaxies]{\bf {\sl ROSAT\/} and {\sl 
 ASCA\/} Observations of X-ray Luminous 
starburst Galaxies : NGC3310 and NGC3690}

\author[A. L. Zezas,I. Georgantopoulos and M. J. Ward]
{A.L. Zezas$^1$, I. Georgantopoulos$^2$ and  M. J. Ward$^1$ \\
$^1$ Department of Physics and Astronomy, University of Leicester, 
Leicester, LE1 7RH \\
$^2$ National Observatory of Athens, Lofos Koufou, Palaia Penteli, 
15236, Athens, Greece } 

\maketitle
\begin{abstract}
We present ROSAT (HRI and PSPC) and ASCA observations of the two 
luminous ($L_x \sim 10^{41-42}$ $\rm erg~s^{-1}$) star-forming galaxies 
NGC3310 and NGC3690. 
 The HRI shows clearly that the sources are extended with 
  the X-ray emission in 
 NGC3690 coming  from at least three regions.  
The combined 0.1-10 keV spectrum of NGC3310 can be described by two components,
 a Raymond-Smith plasma with temperature $\rm kT=0.81^{+0.09}_{-0.12}$ 
 keV and a hard
 power-law, $\Gamma=1.44^{+0.20}_{-0.11}$,  (or alternatively 
 a harder Raymond-Smith plasma with $\rm kT\sim 15$ keV), 
 while there is no substantial excess absorption 
 above the Galactic. The soft component emission is probably due to 
 a super-wind while the nature of the  hard emission is more uncertain 
 with likely origins,  
 X-ray binaries, inverse Compton scattering of IR photons, an AGN or 
 a very hot gas component ($\sim 10^8$ K).  
The spectrum of NGC3690 is similar,  with 
$\rm kT=0.83^{+0.02}_{-0.04}$ keV and 
 $\Gamma=1.56^{+0.11}_{-0.11}$. 
 We also employ more complicated models 
such as a multi-temperature thermal plasma, a non-equilibrium 
ionization code or the addition of a third softer component  
which improve the fit but not at a statistically significant level ($<2\sigma$).  
These results are similar to recent results on the 
archetypal star-forming galaxies M82 and NGC253.

\end{abstract}

\newpage

\section{INTRODUCTION}
 Significant X-ray emission from star-forming  galaxies has been 
known since
the \einstein epoch  (eg Stewart et al. 1982).      
Their X-ray  spectrum in the 0.3-3.5 keV band can be 
 fitted by a thermal plasma of
temperature about 2 keV with an absorption often higher than the Galactic value (eg.
Fabbiano 1988, Kim \etal 1992). However, simple spectral fits  
suggested that the emission mechanism may be more complicated than
a single temperature thermal plasma. 
 Indeed, for a few nearby star-forming galaxies (eg M82 and NGC253) several 
discrete sources, mostly X-ray binaries, are also resolved 
(Long \& Van Speybroeck 1983).  
The \einstein HRI provided  the first evidence for  extended X-ray features 
(eg Watson \etal ,1984, Fabbiano and Trinchieri 1984).
 
 The effective area of \rosat PSPC provided the opportunity 
to dramatically increase the number of starforming galaxies studied in 
X-rays. 
 Hot gas halos and outflows, have also been observed, 
 around many starforming galaxies (eg. Read, Ponman \& Wolstencroft 
 1995, Della Ceca, Griffiths \& Heckman 1996, Read, Ponman \& Strickland 
  1997), 
 as was anticipated based on previous
optical (Heckman \etal 1990) and  theoretical work (e.g. MacLow and Mc
Cray 1988).
 This wind can be produced by
the supply of mechanical energy via stellar winds from evolved
massive stars and  numerous supernovae. The hot gas is accelerated
outwards forming an expanding superbubble, which on reaching the
extent of the minor axis of the galaxy, blows out following  the onset
of the Rayleigh-Taylor instability. 
These outflows can extend for several kpc along the minor axis of the
galaxy, and have major influence on its subsequent evolution. 
 This is especially so in the case of dwarf galaxies, where because of
 their weak gravitational potential, the wind can remove the
 interstellar medium causing star formation to cease,
 (eg Dekel \& Silk 1986, Heckman et al. 1995). 
 Combined ROSAT and ASCA observations of star-forming galaxies, 
 (eg Serlemitsos, Ptak \& Yaqoob 1996) 
  suggest that the soft X-ray band can be fitted with
 a thermal plasma and a power-law model. The plasma temperature is
 $\sim0.6-0.8 \rm{keV}$, with an absorbing column $N_H>10^{21}$ $\rm cm^{-2}$, 
 well above the Galactic value, 
 and the power-law photon index is $\sim 1.7$. 
 The hard X-ray band can be fitted equally well with a high temperature 
 thermal plasma model, $\rm kT >6 \rm{keV}$, (cf. Ohashi et al. 1990).
The soft emission may emanate from the super-wind while the 
origin of the  harder power-law X-ray emission may be 
 due to either X-ray binaries or Inverse Compton scattering of 
infrared-photons by the relativistic electrons generated by the supernovae
 (eg Rieke et al. 1980). The presence of very hot gas 
 ($\sim \rm 10^8 K$),
 or a weak active nucleus cannot be ruled out.     
 However, medium spectral resolution, wide energy band (0.5-10 keV)
 observations of NGC253 and M82 with ASCA, show that the
 above two component 
 models do not give an acceptable description of the
 data (Ptak et al 1997, Moran \& Lehnert 1997). Even the 
 introduction of a third  thermal component ($\rm kT\sim 0.3 keV$) 
 still provides a poor fit to the high-resolution ASCA SIS spectrum 
 (Moran \& Lehnert 1997). This
 suggests that the emission mechanisms are more complex than these 
 simple models can explain.  
  Since the X-ray spectra of Galactic supernova remnants  
 (eg Hughes \& Singh 1994) cannot be well-fit 
 by a simple Raymond-Smith model (their emission is dominated by 
 non-equilibrium processes) it is likely that we are dealing with 
  a similar situation in starforming galaxies. 

 One of the interesting results of the \rosat All-Sky Survey was the
discovery of  star-forming galaxies with X-ray luminosities  $\sim10^{42}
\,\rm{ergs^{-1}} $, approaching those of Seyfert galaxies
(eg Moran et al 1995). 
 In order to gain further insight into the nature of this powerful X-ray
emission and to compare starburst 
 properties with those of the archetypal 
 (less luminous) star-forming galaxies, we have cross-correlated the
\rosat  All Sky Survey Bright Source Catalogue, RASSBSC, (Voges et al. 1996)
 with the spectral atlas of nearby galaxies of Ho, Filippenko and 
 Sargent (1995). 
 The advantage of the latter sample is that it has high quality optical
 spectra
 and thus provides us with a reliable classification for each galaxy. 
 The sample of Ho \etal contains galaxies with declination $\delta> 0^{o}$
 and magnitude $B<12.5$ while the RASS   
 contains almost 20,000 sources over the whole sky.  
The cross-correlation yielded 43 X-ray counterparts  within 1 arcmin 
of the optical galaxy, whereas less than one coincidence 
is expected by chance.  
 Within this sample of galaxies the vast majority are AGNs (Seyferts
$1\&2$, and LINERS), but there are also some star-forming galaxies and  
high luminosity early-type galaxies. Further details of the 
properties of the resulting X-ray sample are given elsewhere (Roberts,
1998).   
The star-forming galaxies  are listed in table 1. 
In this paper  we present X-ray imaging and spectral analysis
for the two most luminous
star-forming galaxies in the sample: NGC3310 and NGC3690, excluding
NGC5905 which underwent some form of transient phenomenon as yet
not understood.
Both \rosat and
\asca observations are available in the HEASARC archive for these 
two galaxies.

 We use a value of $\rm H_o=50 km~s^{-1}~Mpc^{-1}$ for the Hubble constant,
 throughout this paper. All errors quoted refer to the 90 per cent 
 confidence level.

 \begin{table}
\begin{center}
\begin{minipage}{6.5cm}
 \caption{The X-ray luminous starburst galaxies in the Ho \etal sample}
 \begin{tabular}{cc}
Galaxy & $\rm{X-ray\ luminosity^{\dag}}$\\ 
M82 & 40.88 \\
NGC3310 & 40.89 \\
NGC3690 & 41.62 \\
NGC4449 & 39.24 \\
NGC5204 & 39.54 \\
NGC5905 & $42.18^{\ddag}$ \\
 \end{tabular}
\\\ 
\small{$\dag$ This is the log of the X-ray luminosity in the ROSAT band,
from the RASS data.}\\
 \small{$\ddag$ This galaxy has show significant long time variability
(Bade \etal 1996); This luminosity refers at the outburst epoch}.\\
 \end{minipage}
\end{center}
\end{table}

\subsection{  NGC 3310}

    NGC3310 is a well studied nearby galaxy of Sbc(r)pec type (Mulder
\etal 1996). Its recession velocity is $980 \rm{km \ s^{-1}}$  
(de Vaucouleurs et al. 1991)  
which implies  a distance of 19.6 Mpc.
 Its optical image shows some interesting features. The most
prominent is an arc and bow structure which extends for $100''$ to the
North West of the centre of the galaxy. Bertolla and Sharp (1984) 
propose that the
arc could be part of a spiral arm and the bow is the remnant of an
old jet.  Balick and Heckman (1985) suggest that the whole structure
is the remnant of a collision between the galaxy and a dwarf
companion.  This is in agreement with the suggestion of Mulder \etal
(1985) who characterize the system as a young merger based on the
anomalies found in its rotation curve. It should be noted that
in the radio and near-infrared images we  do not see two
resolved  nuclei  (Telesco and Gatley 1984, Balick and Heckman
1981). This could mean that the merger phase is now complete as
suggested by Balick and Heckman. 

 Another interesting morphological feature is a giant HII
region situated $\sim12''$ SW  of the nucleus (Balick and Heckman 1981).  
 Its size is comparable to the largest
extragalactic HII regions known, and its spectral  characteristics
are dominated by signatures of  Wolf Rayet stars (Pastoriza \etal
1993).

 The starburst nature of the activity has been confirmed by numerous
observations in many spectral bands (e.g. Telesco and Gatley 1984,
Smith \etal 1996, Balick and Heckman 1981). Evolutionary
synthesis models of the starburst give an estimated age range of $10^{7}$ to
$10^{8}$ years, (Van der Kruit and de Bruyn 1976, Balick and Heckman
1981, Telesco and Gatley 1984, Pastoriza \etal 1993).

 NGC3310 has been previously observed with the Imaging Proportional
Counter (IPC) on board the \einstein observatory
 (Fabianno et al. 1992) yielding  
an X-ray flux $\rm f_x =1.1 \times 10^{-12} $   $\rm erg~s^{-1}~cm^{-2}$
in the 0.3-3.5 keV band for a 5keV bremsstrahlung model and 
Galactic absorption. 
This corresponds to an X-ray luminosity of 
 $5.0\times10^{40}\,\rm{erg\,s^{-1}\,cm^{-2}}$.

 \subsection{ NGC 3690}
  
 NGC3690 is also a merging system, but in contrast to NGC3310 it is in
an earlier merger phase. The system is composed  of
NGC3690, which is the western part of the merger and IC694,
the eastern component. The whole system is also known as Arp 
299 or Mrk 171. Casoli \etal (1989) propose that there is a third component in
the system thus forming an interacting triplet.
Its  recession velocity is 3159 $\rm km~s^{-1}$ (Sanders and Mirabel 1985), 
implying a distance of 63.2 Mpc. 
At that distance the projected separation
of $22.5''$ of the two components  corresponds to 6.9 kpc.
 The infrared surface brightness of NGC3690 is at least twice that of IC694 (Friedman \etal  1987). The
total mass of gas  calculated  from radio observations
is about $2\times 10^{11}M_{\odot}$ (Casoli \etal. 1989) 

 The high IR luminosity$(L_{FIR}=1.2\times10^{12}L_{\odot})$
(Soifer \etal 1987), results from re-radiating dust heated by the starburst
activity. This  is
supported by multi-wavelength  studies (Gehrz \etal 1983,
Friedman \etal 1987, Nakagawa \etal 1989). Detailed analysis of
emission line ratios using the diagnostic diagrams of
Veilleux and Osterbrock (1987) clearly indicate a stellar origin 
of the ionizing continuum (Friedman \etal 1987).
 The infrared  emission is extended over a region of several kpc,
 but the main sources are located close to the two nuclei (Friedman
\etal 1987). Modeling of the starburst has been carried out  by Gehrz \etal
(1983) and Nakagawa \etal (1989). The latter find a starburst age of
about 10Myrs. An interesting point is that the two nuclei 
have different properties, implying either a
different age and/or a different Initial Mass
Function (IMF) for each starburst event. 

\begin{table*}
\caption{Summary of the observations}
\begin{center}
\begin{tabular}{cccccc}

Galaxy & Satellite & Instrument & Date Observation started& Exposure
  Time\dag \\
 &  & & & Ksec \\
 NGC3310 & \rosat & PSPC & 1991-11-16 & 9.114  \\
  & \rosat & HRI & 1995-04-17  & 41.842 \\
  & \rosat & HRI & 1994-11-23 & 4095  \\
  & \asca & & 1994-04-17 & GIS: 10.4, SIS:10.4 \\
  & \asca & & 1994-11-13 & GIS:11.6, SIS:10.2 \\
 NGC3690 & \rosat & PSPC &1993-04-22 & 3.534 &  \\
  & \rosat & PSPC & 1991-11-18 & 6.391 \\
  & \rosat & HRI & 1993-04-18 & 6.751  \\
  & \asca & &1994-04-06 & GIS:5.7, SIS:5.3\\
  & \asca & & 1994-12-01 & GIS:38.2, SIS:35.5 \\
\end{tabular}
\end{center}

\end{table*}

 NGC3690 has been previously observed with the High Resolution Imager
(HRI)  on
board the \einstein observatory. 17 counts were detected implying an
X-ray flux $f_{x}=4.8\times10^{-13}\rm{ergs^{-1}cm^{-2}}$ in the 0.2-4.0keV
band, for a 5 keV bremsstrahlung model and Galactic absorption
(Fabianno et al 1992). This corresponds to an X-ray luminosity of
$2.3\times10^{41}\,\rm{erg\,s^{-1}\,cm^{-2}}$.

\section{OBSERVATIONS AND DATA REDUCTION}

\subsection{  The {\sl ROSAT\/}~PSPC Observations}
  
NGC3310 and NGC3690 were each observed on two occasions  with the Position
Sensitive Proportional Counter (PSPC) (Pfefferman \etal 1987)
on board \rosat (Tr\"{u}mper \etal 1984).  The details of each observation are
 given in table 2.

 For the data reduction we have followed the standard procedure, using
the \textit{ASTERIX} package. We  excluded those data with Master Veto 
rates higher than 170cps.  We extracted a PSPC spectral image
(channels 11 to 201) with a pixel size of $15''$. To obtain an X-ray
spectrum we extracted data from a circular region
 of $1.5'$ radius around the X-ray centroid. The background was estimated
from an annular region between radii of $15'$ and $8.8'$ from the centroid, 
after exclusion of all the discrete sources
detected with the PSS algorithm (Allen 1992) down to the $5\sigma$ level. 

\subsection{  The {\sl ROSAT\/}~HRI Observations}
 Both galaxies  have been observed with the High
Resolution Imager (HRI) (David \etal 1997)  on board \rosat. 
 HRI  covers a field of $38'$ diameter and it has a
spatial resolution of about $5''$ (the FWHM of the XRT+HRI Point Spread
Function). We note that the HRI has very limited spectral
resolution so we cannot use it 
for spectral analysis (David \etal 1997).
 The screening  of the data has been carried out  using the 
 \textit{ASTERIX} package.
 We have rejected all data with aspect errors greater than 2. 

\subsection{ The {\sl ASCA\/}~Observations}

 Both galaxies have been observed with the \asca satellite (Tanaka
\etal 1994). 
On board \asca there are four instruments: two Gas Imaging
Spectrometers (GIS2 and GIS3) (Ohashi \etal 1996)  and two Solid-State Imaging
Spectrometers (SIS0 and SIS1) (Gendreau 1995). 
 For screening of the data we have followed the standard procedure
 and used the {\textit ASCACLEAN} program in the 
FTOOLS package with the parameters described in the ABC \asca 
 Guide (Yaqoob \etal 1997). After the standard processing we inspected the
light curves and removed time intervals with unusually high numbers
of counts, which may result from background particle contamination.
 The extraction of the source and the background spectra was 
been carried out using the XSELECT program within the FTOOLS package. 
 We extract the source spectrum using a circular region of 
 2.7 and 5.5 arcmin radius for the SIS and GIS respectively.  
For the background regions 
we have selected ten source free zones, each of area about 6 and 10 
 $\rm arcmin^2$ for the SIS and GIS respectively. 
 Following extraction of the source and background spectra, we ran
the SISrmg program to create the SIS response matrices for each SIS
source spectral file. For the GIS spectral files we used the
$\rm{gis2v4\_0.rmf}$  and $\rm{gis3v4\_0.rmf}$ for GIS2 and GIS3 respectively.
We created the Ancillary Response Matrices for each file by running 
the ASCAarf  programs within the FTOOLS package.
  
 \section{ SPATIAL EXTENT ANALYSIS}

\subsection {NGC3310}
 We used the archival \textit{HRI} observations   
to obtain information on the morphology  of the X-ray emitting regions
 and compared these with images of these galaxies at other wavelengths. 
 First we searched for extended emission by comparing
the background subtracted radial profile of the galaxies with the
radial profile of a point source. For this purpose we used as a reference 
the point source corresponding to 
the star AR-Lac. We retrived archival data for an
on-axis observation of AR-Lac, and applied the same extraction process 
as used for the galaxy. The comparison
of the radial profiles of NGC3310, and the radial profile of
AR-Lac is shown in figure 1.
This figure clearly shows that the X-ray source 
has a radius of $\sim 0.5 $ arcmin (at larger radii 
the emission from the galaxy begins to blend with the profile of AR-Lac, 
within the $2\sigma$ error bars). Although this 
is a somewhat arbitrary point at which to define the extent of the emission,
it gives a lower limit to the source size. At the distance of 
NGC3310 this corresponds to $\sim3$ kpc. 
 We note that fitting a Gaussian function to the radial profile gives a 
 similar source size (FWHM=1.6 arcmin). 

\begin{figure}
\rotatebox{270}{\includegraphics[height=6.5cm]{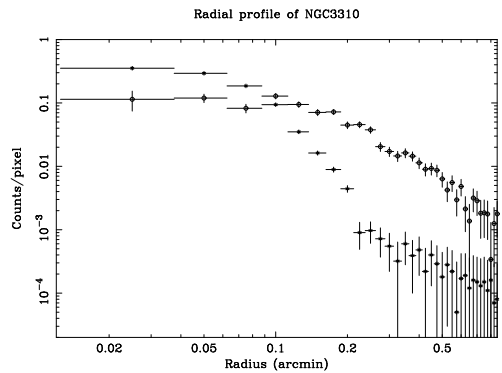}}
\caption{The HRI radial profile of NGC3310 (circles) overlaid on the radial
profile of AR Lac (stars).}
\end{figure} 

 In order to compare the distribution of the X-ray emitting gas with
the optical image we have resampled the X-ray image with a pixel
size of 1.5 arc-seconds and then  smoothed this image using a
two-dimensional gaussian of  FWHM=3.5 arc-seconds (2.3
pixels) following Della Ceca \etal (1996). The final pixel size 
is $5.4''$, equal to the XRT+HRI PSF.
Then we obtained  contours from this image  
corresponding to 2.2, 2.0 ,1.7, 1.33, 0.88, 0.44, 0.35, 0.26, 0.17 
 $\rm{counts\,arcsec^{-2}}$  and  overlaid these onto an optical image
obtained from the  Digitized Sky Survey.      
This image is shown in figure 2. The main problem we faced
in attempting to overlay the two images is that of frame 
registration. There are no X-ray
sources within the HRI image which have an obvious optical
counterpart. Thus in order to overlay the X-ray contour plot onto the
optical image, we are forced to assume that the centroid of the 
X-ray source  corresponds to that of the optical nucleus.
In the X-ray image of NGC3310 we detect two other sources in
addition to the nucleus, at a significance above
$10\sigma$. The details concerning these sources  are presented in table 3. 
 These point sources may correspond to luminous X-ray binaries or
young supernova remnants. Unfortunately there are no identified
optical counterparts for these X-ray sources on the POSS plates.

\begin{figure*}
\rotatebox{270}{\includegraphics[height=15cm]{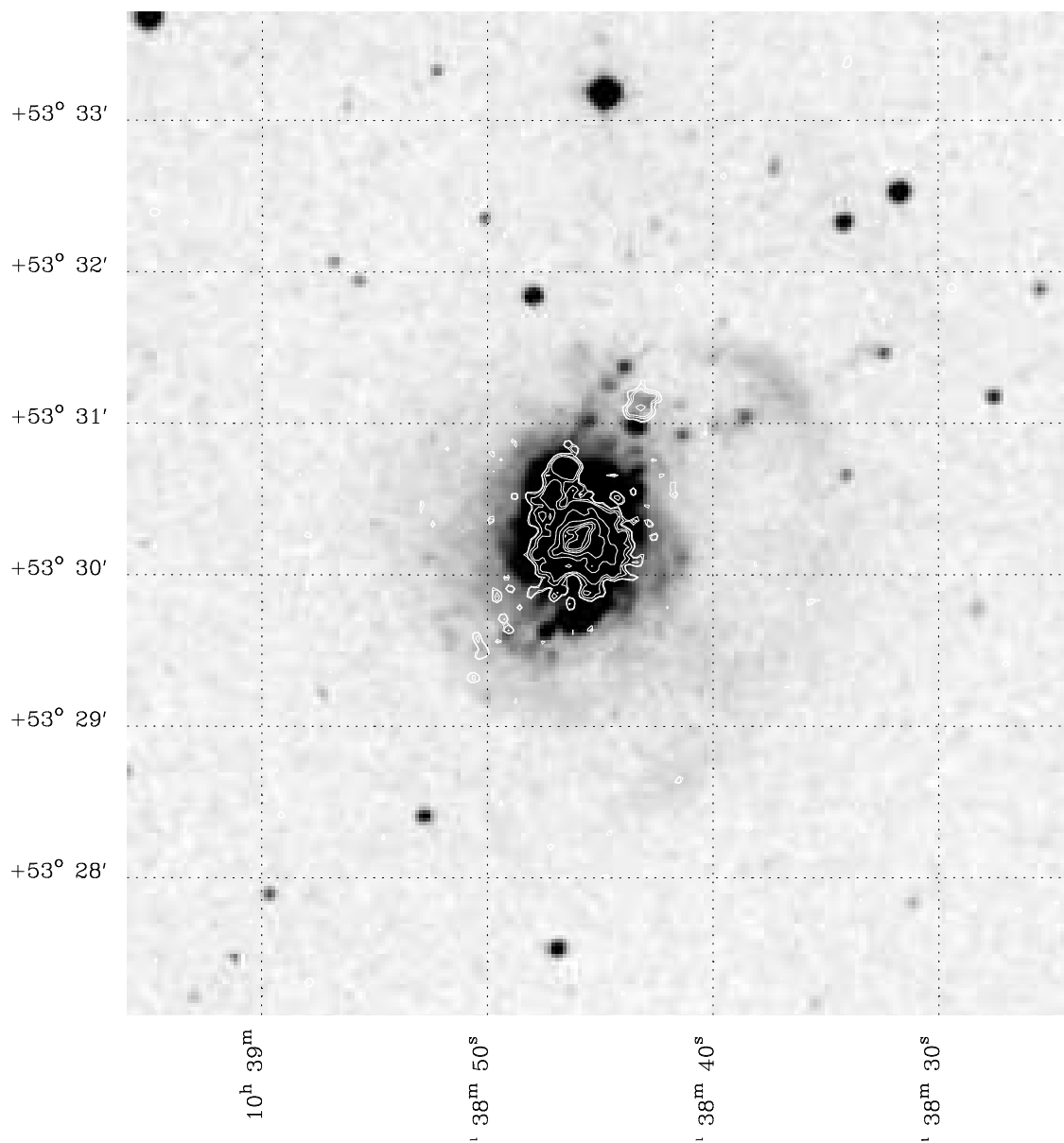}}
\caption{X-ray contours from the HRI observation of NGC3310  overlaid
on a POSS image. The contours are in levels of 2.2, 2.0, 1.7, 1.33,
0.88, 0.44, 0.35, 0.26, 0.17 $\rm{counts\,arcsec^{-2}}$.}
\end{figure*}

\begin{table*}
\caption{Sources in the \textit HRI field of NGC3310}
\begin{center}
\begin{tabular}{ccccccccc}

 &RA (J2000) & & & Dec (J2000) & & Count rate & Flux & Significance\\ 
 10 &38 &43.2 & +53 &31 &07 & 2.0$\pm$0.27$\times10^{-3}$  
  & $2.3\times10^{-14}$ & 15.5 \\
 10 & 38 &46.7 & +53 & 30 & 38 & 2.8$\pm$0.26$\times10^{-3}$  
  & $3.07\times10^{-14}$ & 23.4 \\
\end{tabular}
\end{center}
\end{table*}

\subsection {NGC3690}
   
The HRI analysis of NGC3690 followed the same procedure as
for NGC3310. The image pixels were binned to a size of $1.5''$ and then
smoothed using  a two-dimensional gaussian of FWHM  $3.5''$.
Finally, we overlaid the X-ray contours on a POSS plate image 
 obtained from the Digitized
Sky Survey, figure 3; the contours correspond to
0.57, 0.44, 0.40, 0.35, 0.31, 0.27, 0.22, 0.18, 0.13 and 0.11
  $\rm{counts\,arcsec^{-1}}$ respectively. 
The most striking feature is the existence of three separate
sources in the X-ray
image. The two most luminous sources correspond to the 
 two merging  nuclei. Their X-ray fluxes are given  in table 4.
The third X-ray source corresponds to a hot spot seen in the infrared and
radio images to the north of  NGC3690 (the western component). 
 Apart from these three sources there are marginal detections of other
sources, 
but at a low level of significance (below $3\sigma$), and will not be 
discussed further.
We note that the relative strengths of 
the three main X-ray sources follow quite well the relative strengths
of the near infrared sources, suggesting a possible common origin for the IR 
and the X-ray emission. 

 We have searched for extended emission from the two resolved
nuclei. From figure 3 it is clear that the emission from the two resolved
nuclei is extended and not symmetrical. 
 Fitting a Gaussian function to the 
 radial profiles of NGC3690 and IC694 yields 
a FWHM of 1.5 and 1.8 arcmin respectively. 

 Comparison between the X-ray images and radio (VLA-A observation of the
$\rm{H92\alpha}$ radio recombination line, Zhao \etal, 1997)
 and near-IR images  (Wynn-Williams \etal, 1991), 
 available in the literature, shows some clear similarities.
 As the radio and near-IR emission is clearly associated with the 
 star-forming regions, the spatial coincidence implies  
 that the soft X-ray emission has its origin 
 in the starburst. We note that we do not
see any evidence of X-ray emission from hot connecting the two galaxies, 
unlike the situation in some other interacting 
star-forming galaxies (e.g. the Antennae, Fabbiano \etal 1997).  

\begin{figure*}
\rotatebox{270}{\includegraphics[height=15cm]{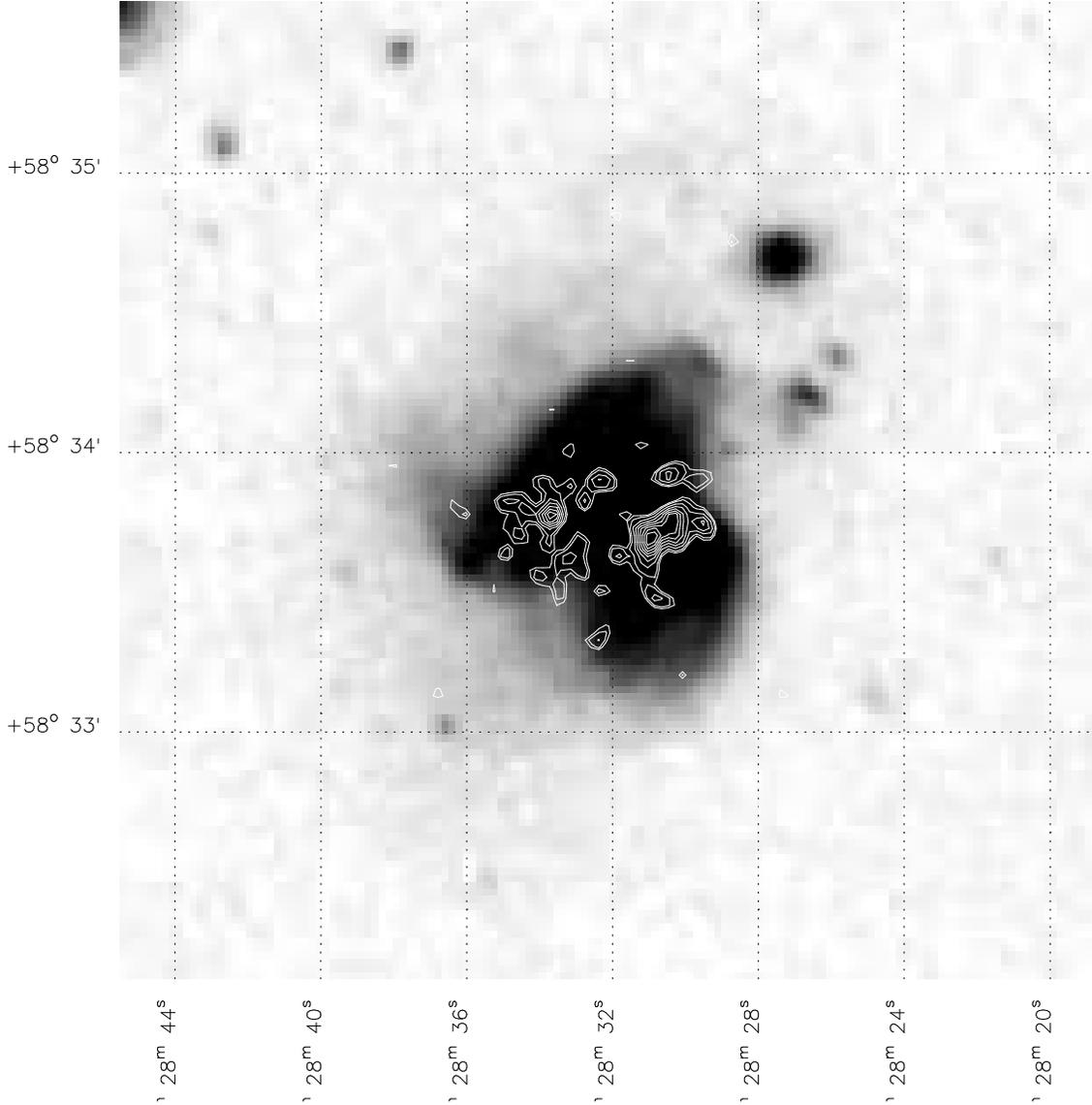}}
\caption{X-ray contours from the HRI observation of NGC3690 overlaid on a POSS
image. The contours are in levels of 0.57, 0.44, 0.4, 0.35, 0.31,
0.26, 0.22, 0.18, 0.13, 0.11 $\rm{counts\,arcsec^{-2}}$.  }
\end{figure*}

\begin{table*}
\caption{Sources in the \textit HRI field of NGC3690}
\begin{tabular}{cccccccc}

 &RA (J2000) & & & Dec (J2000) & & Count rate (cps) & Flux (0.1-2.5keV) \\ 
 11 & 28 & 30.7 & +58 & 33 & 50 & 10$\pm$1.2$\times10^{-3}$ 
   & $1.18\times10^{-13}$  \\
 11 & 28 & 34.1 & +58 & 33 & 51 & 15$\pm$1.5$\times10^{-3}$  
   & $1.72\times10^{-13}$  \\
 11 & 28 & 30.0 & +58 & 34 & 03 &  3$\pm$0.6$\times10^{-3}$ 
    & $3.17\times10^{-14}$ \\
\end{tabular}
\end{table*}

\section{ SPECTRAL ANALYSIS}
 
  We fitted all the datasets simultaneously,
thus covering  an energy range from 0.1 to 8.0 keV. 
We rejected all ROSAT PSPC data points below 0.1 and above 2.0 keV;
and used only the SIS and GIS data in the range 0.6 to 8.0
keV and from 0.7 to 7.0 keV respectively. In order to apply $\chi^{2}$
statistics we re-binned the spectra so as to contain at least 20 counts
per bin.
 We used the software package XSPEC (v10) to obtain the spectral fits.

\subsection{NGC3310}

  We first fitted simple one-component models: 
  a thermal bremsstrahlung model, a power-law model    
  and finally a Raymond-Smith plasma model (Raymond \& Smith 1977).  
 These models were rejected at greater than  the $99.9$ per cent confidence
  level. The results of the spectral fits are given in table 5.  

 We then fitted two component models: a two temperature Raymond-Smith
plasma (ray-ray hereafter) and a Raymond-Smith plasma combined with a
power-law (po-ray), following the results of Moran \& Lehnert (1997) and 
 Ptak et al. (1997) on the starburst galaxies M82 and NGC253. 
These models provided a much better fit than the simple models described
above, at a confidence level over $90$ per cent for an addition of 
two free parameters (eg Bevington \& Robinson 1992). 
 We consider first the po-ray model.  
We obtained a photon index of $\Gamma=1.44^{+0.20}_{-0.11}$ and a temperature
 of $kT=0.81^{+0.09}_{-0.12}$ keV with a reduced $\chi^{2}=1.115$, 
  where the abundance is fixed to the solar value
and the absorbing column density
to the Galactic value, $N_H=0.7\times10^{20}$ $\rm cm^{-2}$, (Stark et al. 1992). 
 Leaving the absorbing column density as a free parameter further improves the
fit (reduced $\chi^{2}=1.031$);  this is significant at a confidence 
 level of $>99$ per cent for one additional parameter. In contrast,  
 leaving the abundance as a free parameter does not improve the fit
 further at a statistically significant level. 

 In the ray-ray case we do not obtain  a  significantly 
 improved fit   
 compared to the po-ray model (reduced $\chi^{2}=1.022$).  
 Figure 4 shows the spectrum of NGC3310 with the double Raymond-Smith model.  
 We obtained temperatures of 
$0.80_{-0.04}^{+0.07}$ keV and $14.98^{+13.52}_{-4.88}$ keV for 
 an absorbing hydrogen column density of
$N_{H}=1.37^{+0.50}_{-0.32}\times10^{-20}\ \rm cm^{-2}$. 
 When we fixed the column density to the Galactic value 
 ($N_H \approx 1.0\times 10^{20}$ $\rm cm^{-2}$)  
  we obtained  $0.84^{+0.05}_{-0.07}$ and $\rm kT\approx 64$ keV.
 We also introduced two different absorbing column densities,  
 for the soft and the hard components, 
 following the analysis of M82 by Moran \& Lehnert (1997) who showed  
 that the hard X-ray component is obscured at the nucleus by 
 $N_H\sim 10^{22}$ $\rm cm^{-2}$. However, we do not 
 find similar evidence here, since the hard component column has a value 
 of $N_H\approx 1.77\times10^{20}\,\rm{cm^{-2}}$. 
  Finally, we attempted to measure the element abundance 
 in order to compare it with that for other star-forming  galaxies 
 (Sansom \etal 1996, Serlemitsos \etal 1996),
 which appear to show a systematic trend towards sub-solar 
 abundances. Unfortunately the data for our galaxies are 
 not able to set any useful constraint on the  abundances.
 
\begin{figure}
\rotatebox{270}{\includegraphics[height=8.0cm]{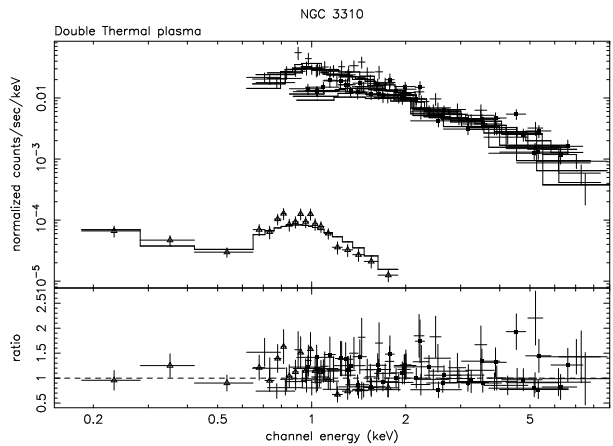}}
\caption{The top panel shows \rosat
and \asca spectra of NGC3310 with the best fit double Raymond Smith
model. The bottom panel shows the ratio of the data points to the model.}
The SIS data are marked with dots while the PSPC data are marked with triangles. 
\end{figure}

We do not detect the  $FeK_{\alpha}$ line at 6.7 keV
as would be expected for a hot thermal plasma. The low signal/noise data above
$\sim5 \rm{keV}$,
does not allow us to set a useful upper limit on the line
equivalent width.
 The lack of a strong FeK line could be explained by lower than solar 
 abundances, as is probably the case in
 other well-studied starbursts such as NGC253, (Ptak \etal 1997), 
 and the Antennae, (Sansom \etal 1996). 

We note that a previous preliminary analysis of ASCA data by Serlemitsos
\etal (1996), find a spewctral slope consistent with our value. However, they
derive a considerably higher column density, but without the benefit of using
the lower energy \rosat data.

\begin{table*}
\caption{Spectral fitting results for NGC 3310}
\begin{tabular}{cccccc}
Parameter &  Power-law & Single Temperature & Double Temperature &
Raymond Smith + & \\
          &  & Raymond-Smith & Raymond Smith &
Power-law \\
kT (KeV) &  & $7.35^{+1.24}_{-1.18}$
& $0.80^{+0.07}_{-0.04}$  & $0.81^{+0.09}_{-0.12}$ &
\\
 &  & & $14.98^{+13.52}_{-4.88} $ & & \\
\g  &  $1.68^{+0.08}_{-0.08}$ & &  & $1.44^{+0.20}_{-0.11}$  \\
$N_{H}(10^{20}cm^{-20})$  & $ 3.15^{+0.63}_{-0.52}$
& $1.90^{+0.45}_{-0.36}$ & $1.37^{+0.50}_{-0.32}$ & $1.74^{+0.68}_{-0.40}$ \\ 
\x2 / d.o.f. & $222.9/175$ & $239.3/175$ & $168.7/165$ & $170.2/165$   \\  
Flux$^{\dag}$ (0.1-2.0keV)  & $0.95$  & $0.86$  & $0.9$  & $1.09$ \\
Flux$^{\dag}$ (2.0-10.0keV)  & $1.88$ & $2.12$ & $2.10$ & $2.10$ \\
Total Luminosity$^{\ddag}$ (0.1-10keV) & 1.3 & 1.4 & 1.4 & 1.5 \\
\end{tabular}
\\
\small{\dag All the fluxes are in units of $10^{-12}\,\rm{erg\,s^{-1}\,cm^{-2}}$}.
\small{\ddag The luminosity is in units of $10^{41}\,rm{erg\,s^{-1}}$.} 
\end{table*}

\subsection{NGC3690}

The spectral analysis follows the same procedures as for NGC3310. 
 Simple one-component models such as a power-law and a Raymond-Smith 
 plasma are similarly rejected (see table 6). 
  Two-component models, such as a 
 po-ray and a ray-ray model 
 (with the hydrogen column density as a free component) 
 give  $\rm{\Gamma=1.56^{+0.11}_{-0.11}}$,
 $\rm kT=0.83^{+0.02}_{-0.04}$, $N_H=2.4^{+0.6}_{-0.5}\times 10^{20}$ 
 $\rm cm^{-2}$ and   
 $\rm{kT_{1}=0.83_{-0.03}^{+0.03}}$,
 $\rm{kT_{2}=10.3^{+5.95}_{-2.44}}$, $N_H=1.6^{+0.4}_{-0.4}\times 10^{20}$ 
 $\rm cm^{-2}$ respectively (the Galactic absorption is 
  $N_H=1\times10^{20}\rm cm^{-2}$). 
 In the above model the abundance was 
 fixed to solar. Also, as for NGC3310 when the abundance is a free
 parameter, it cannot be usefully constrained.  
 Figure 5  shows the \rosat and \asca
spectrum of NGC3690 with the best fit double temperature Raymond-Smith
thermal plasma model. We can clearly see from the spectrum 
that there is no evidence for an Fe line at 
 $\sim6.7$ keV. The upper 
 limit to the Fe line equivalent width is $\sim860 \rm{eV}$.
 We also tried to fit the models using a different absorbing 
 column for the hard component. We obtain
$N_{H}=1.7_{-0.8}^{+0.1}\times10^{20}\,\rm{cm^{-2}}$, which is 
comparable to the previous best fit values,  
 while the $\chi^2$ is not significantly improved for 
 one additional parameter. Thus there is no evidence for 
 excess absorption in the hard component, in contrast to the 
results for M82 (Lehnert \& Moran 1997).  
  We also note that both the ray-ray and the po-ray models  do not provide 
 an adequate fit to the data as they can be rejected at over 
 the 99 per cent level of confidence using a $\chi^2$ goodness of fit.

\begin{figure}
\rotatebox{270}{\includegraphics[height=8.0cm]{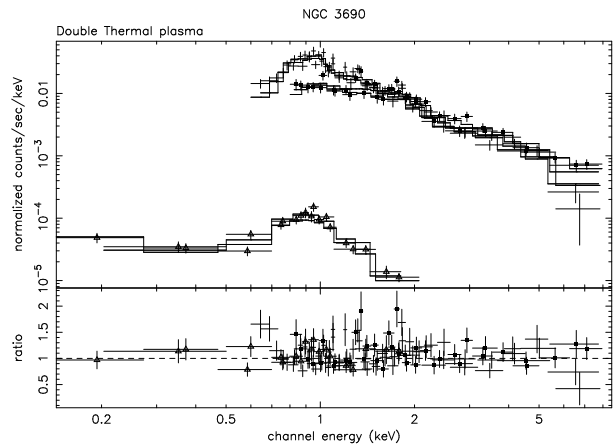}}
\caption{The top panel shows \rosat
and \asca spectra of NGC3690 with the best fit double Raymond Smith
model. The bottom panel shows the ratio of the data points to the model.}
The SIS data are marked with dots, while the PSPC data are marked with
triangles.
\end{figure}

We therefore attempted to fit more complicated models, such
 as multi-temperature components, and a power-law
 distribution of temperatures. This is  model `cevmekl' in XSPEC, (see Done
 and Osborne 1997). This model gives a best fit of reduced
 $\chi^{2}=1.286$, with a maximum temperature of $\sim57$ keV, a 
 power-law index of the temperature distribution of
$\alpha=0.22^{+0.11}_{-0.10}$
and $N_H=3.21^{+0.73}_{-0.48} \times 10^{20}$ $\rm cm^{-2}$.
 A combination of a power-law and the 
multi-temperature model gives a slightly better fit  (reduced
$\chi^{2}=1.211$) with  $\Gamma=1.52^{+0.13}_{-0.12}$, 
 $\rm{T_{max}=1.17^{+0.25}_{-0.16}}$ and
$\alpha$ fixed to 1. 
Finally, we considered using a non-equilibrium-ionization
 model (NEI), such as that applied to hot gas resulting from
 supernova explosions (Hughes and Singh 1994). This model assumes
 that the gas is instantaneously shock heated, and then applies
 corrections to the Raymond-Smith thermal plasma code to account for
 the non-equilibrium ionization fraction. An additional variable
 in the NEI model is {\it nt}, which is the product of the electron density
 and the time elapsed since the passage of the shock wave. In principle
 the results from fitting NEI models can be used to estimate 
 the characteristic timescale over which the gas
 reaches ionization equilibrium. Clearly for a situation
 in which multiple supernova occur over an extended time 
 period, it will not be possible to identify a unique timescale. 
 However, NEI models have been applied to the optically thin plasma
 in galactic scale superwinds, such as observed in M82, Tsuru et al
 (1998). Unfortunately in the case of NGC3690, we do not have the 
 benefit of measurements of the low energy lines of silicon and 
 magnesium needed to quantify the importance of non-equilibrium 
 effects. For NGC3690 the NEI models give a best fit with a combination of a
power-law and a NEI model. However the reduction in the \x2 is not
statisticaly significant at a level above 90\%,  
and so we conclude that our data does not justify further consideration
of the NEI models. However, if new soft X-ray emission line data
become available, it would be worthwhile returning to this question.

\section{DISCUSSION}

\subsection{The soft X-ray  emission}

The HRI images clearly show the extended X-ray emission 
in both NGC3310 and NGC3690/IC694. This  supports a thermal origin 
for the soft emission, which 
may be due to a galactic scale super-wind eg Heckman et al. (1993). 
 Such a phenomenon
is observed in other star-forming galaxies, M82 (Strickland,
Ponman \& Stevens 1997), NGC1569 (Della Ceca et al.  1996) and
NGC4449 (Della Ceca et al.  1997). 
According to this scenario, supernova in the starburst region create a hot 
($\sim 10^8$ K) bubble. The dense shell around the bubble 
fragments due to Rayleigh-Taylor instabilities and forms 
the optical emission line filaments and is probably a source of 
 soft X-ray emission. The optical data also support this model. 
 In the case of NGC3690 the $H_{\alpha}$ images 
show plumes and filaments extending out to several kpc (Armus, Heckman \&
Miley 1990). 
 Additional evidence comes from the extended synchrotron radio emission 
which may arise from the electrons 
 produced by supernova remnants in the starburst regions, (Gehrz \etal
1983).
In the case of NGC3310 the situation is more complicated 
as its low inclination angle ($\theta\approx 32^\circ$) makes 
detection of any outflow along the galaxy's minor axis very difficult. 

In order to further test the validity of the above model, we can estimate the 
expected X-ray luminosity of the super-wind and compare 
this with the observed luminosity of the soft Raymond-Smith component. 
 First we relate the X-ray
luminosity of the gas contained in a supperbubble with the basic
properties of the starburst; its bolometric luminosity and
age. Heckman \etal (1996) have calculated the total X-ray luminosity
from a superbubble by integrating over its volume,  and using the
expressions for the gas density and the bubble radius derived by MacLow
and McCray (1988). They find that   
$\rm{L_{X} \simeq \
5\times10^{39}L_{41}^{33/35}n^{17/35}t_{7}^{19/35}\ erg\ s^{-1}}$,
where $L_{41}$ is the injection rate of mechanical energy into the
bubble  in units of $10^{41} \rm{erg\,s^{-1}}$,  n is
the gas density and $t_{7}$ is the 
age of the bubble in units of 10Myrs.
 Considering that the formation of the bubble begins, as a result of
 strong winds from O stars
following the starburst, we can assume that the age of the bubble is almost the same as
the age of the starburst. 
Using the relation  
$\rm{dE/dt=7\times10^{42}L_{IR,11}\ erg\ s^{-1}}$, (Heckman et al. 1996) 
which correlates the total infrared luminosity (in units of $10^{11}
\rm{L_{\odot}}$)  of a starburst with age greater than
$10^{7}\,\rm{yr}$ with the kinetic energy deposition
rate from supernovae and stellar winds, we
can estimate the X-ray luminosity. Using the IRAS data from 
Soifer \etal (1987), the IR luminosity of NGC3310
is $5.5\times10^{10} \rm{L_{\odot}}$. Assuming a gas density of $\sim 1
\rm{cm^{-3}}$ and a typical starburst age of 10Myr we obtain  an X-ray
luminosity $L_{x}=1.5\times10^{41} \rm{erg \ s^{-1}}$ in the ROSAT
band. This luminosity is larger than the measured luminosity in
the same energy band of $(L_X=4.7\times10^{40}$ $\rm erg~s^{-1})$, but 
considering the large uncertainties in the model parameters involved,
we conclude  that the super-wind model remains a possible
scenario for the origin of the soft X-ray emission. 
In NGC3690/IC694 the situation is more complex, as 
 there are at least three distinct starburst components 
(Nakagawa et al. 1989) which are unresolved by IRAS. 
 In order to obtain an estimate of the  mechanical energy  input for each component,
 we used the ground based mid-IR data (10-32 microns) 
of Wynn-Williams et al. (1991). These observations have sufficient 
 spatial resolution to separate the three components. 
 We then make the reasonable assumption that  
 the fractions of the total IRAS flux which arise from  each 
 of the three star-forming 
 components, are similar to the fractions in the  mid-IR band. 
 Using the same formulae (Heckman et al. 1996) as in the analysis of
 NGC3310 above,  
 and assuming a starburst age of 10 Myr and a density of 
 $n=1\,\rm{cm^{-3}}$, we predict  $L_x=1.3\times 10^{42}$,  $1.1\times10^{42}$ 
 and $3.2\times 10^{41}$  $\rm erg~s^{-1}$ for the A,B and C
components respectively (following notation of Gehrz et al., 1983).
 Again the predicted 
luminosity is  higher than the measured soft
(0.1-2.5keV) X-ray luminosity ($4.7\times 10^{41}\rm erg~s^{-1}$).
This could possibly result from the fact that some fraction of the
far-infrared emission measured by IRAS, may arise from regions more
extended than those producing the X-rays. Another possibility is that
an additional source of heating may be present, perhaps due to stars
of later spectral type than those producing the winds.

The soft X-ray spectra of NGC3310 and NGC3690  are very similar. 
 A  Raymond-Smith component 
with a temperature of $\rm kT\sim 0.8$ keV dominates the spectrum at 
soft energies and could originate from the super-wind. 
 This temperature is similar to that found for 
 other star-forming galaxies, eg  M82 and NGC253 (Ptak et al. 1997), 
 NGC1569 (Della Ceca et al. 1996), NGC4449 (Della Ceca et al. 1997)  
 NGC6240 and NGC2782 (Schulz et al. 1997).
 In the case of NGC3690 where the photon statistics are good, 
the SIS spectral fit yields a poor $\chi^2$ fit. This could be
suggestive of a hot tenuous gas component. Indeed, if the expanding gas  
has not had time to reach thermal equilibrium, its spectrum will not 
be well represented  by the  Raymond-Smith model.
The use of a non-equilibrium code by Hughes et al. improved the fit, 
but again the $\chi^2$ is not accepted at the 98 per cent 
confidence level. This is not surprising as 
even a non-equilibrium model may be inadequate to fit an 
ensemble of supernova remnants occuring in several star-forming regions
as is the case for NGC3690/IC694.

\subsection{The hard  X-ray emission}
 The spectrum at X-ray hard energies can be represented, in both galaxies,
 by a power-law ($\Gamma \sim 1.4-1.6$), or alternatively by a 
 Raymond-Smith component with high temperature, (kT$>$10 keV).
 There is no strong evidence for the presence of a large amount of
 obscuration as is
 the case for NGC253 and M82 (Ptak et al. 1997). 
 There are at least four possibilities for the origin 
 of the hard X-ray emission: a) Inverse Compton scattering of 
 the IR photons produced in the starburst by electrons 
 from the numerous supernova remnants (eg Schaaf et al. 1989) 
 b) emission from 
 a low-luminosity AGN  as is the case in NGC3628 (Yaqoob \etal 1995) 
   c) thermal emission from a very hot gas ($T\sim 10^8$ K)
  and d) emission from  X-ray binaries (Griffiths \& Padovani 1990). 
 We discuss each of these possibility in turn. 
We consider first the possibility, that the hard X-ray emission
arises from Inverse Compton (IC) scattering of the copious infrared photons
off of the 
relativistic electrons generated by the numerous supernovae. 
Support for this, especially in the case of NGC3310, comes from the
similarity (within the errors) of the spectral index of the hard
X-rays and the spectral index of the radio emission
($\alpha_{rad}=0.61^{+0.03}_{-0.03}$ Niklas \etal 1997). 
 Following Schaaf \etal (1989) we can estimate the X-ray luminosity
from Inverse Compton scattering. Unfortunately, we can perform this 
calculation only  for
NGC3310, as there are no suitable radio data available for
NGC3690. From Vall\'{e}e (1993) we have that the ``minimum energy''
magnetic field (see Longair 1992) 
is $\rm{B=0.47\times10^{-5}\,Gauss}$. Then
the minimum energy density is
$\rm{u=(7/3)(B^{2}/8\pi)~erg~cm^{-3} =
2.05\times10^{-12}~erg~cm^{-3}}$.
 Following Schaaf \etal (1989) we have that
$\rm{L_{IC}=(1/3)\sigma_{T}R_{IC}L_{IR}\frac{\epsilon_{e}}{mc^{2}}
\,\gamma_{2}^{0.8}\gamma^{0.2}_{1}}$,
where $\rm{\sigma_{T}}$, $\rm{R_{IC}}$, $\rm{L_{IR}}$ and $\rm{mc^{2}}$,
 $\epsilon_{e}$   
 are the
Thomson cross-section, the thickness of the disk, the far-IR
luminosity, the rest mass of the electron and the energy density 
 of the relativistic electrons, respectively. For a typical galactic
disk $R_{IC}\sim1 \rm{kpc}$ (we cannot measure the actual thickness of
the disk since NGC3310 is almost face on) and for a typical value of the low 
frequency cut-off $\rm{\nu=0.01GHz}$ we have $\rm{\gamma_{1}=150}$
for the lower limit of the Lorentz factor, and $\rm{\gamma_{2}=10^{3}}$
for the maximum Lorentz factor, in order to have IC emission at $~10 \rm{keV}$
(Schaaf \etal 1989).
 So for the Inverse Compton X-ray luminosity we predict a value of
$\rm{2.5\times10^{38}~erg~s^{-1}}$, much lower than the detected hard X-ray
luminosity of NGC3310, thus implying that IC can be only a minor component
of the total X-ray emission from this galaxy. We caution that this
result is quiet uncertain as the calculation depends on parameters
like the volume of the source and the thickness of the disk which are
poorly known. 
   Another possible origin of the hard X-ray emission 
 is the presence of a low luminosity AGN, although there is no evidence 
 for non-starforming nuclear activity from 
 diagnostic emission line ratio diagrams based on optical and near infrared
 spectra.
   
 However, the hard X-ray  power-law spectrum is nevertheless consistent
 with the presence of an active nucleus. In order to test this possibility 
 further we use the $L_{X}/L_{H_{\alpha}}$
 relation from Elvis \etal (1984), where $L_{H_{\alpha}}$ 
 is the luminosity of the broad
 component of the $H_{\alpha}$ line and $\rm L_x$ is the hard X-ray luminosity 
 (2-10 keV). 
 $L_{X}/L_{H_{\alpha}}\simeq40 $, for low-luminosity AGN and thus 
 we estimate 
 $\rm f(H_{\alpha})\sim 5.3 \times10^{-14} \ \rm{erg\ s^{-1} \ cm^{-2}}$ 
 and $2.7\times10^{-14} \ \rm{erg \ s^{-1} \ cm^{-2}}$
 for NGC3310 and NGC3690 respectively.
 A broad component of this strength is easily detectable, but is not seen 
 (Ho et al. 1997).
 The absorbing columns found from the X-ray spectral fitting  (ray-po models
 in tables 5 and 6)  imply low 
 extinctions,  $A_V\sim0.1$ mag for both galaxies (Bohlin et al. 1978).
 Hence a BLR reddended by this amount would still be observed.
 
 The next possibly is the presence of a very hot thermal component. 
A Raymond-Smith model with kT$>$10 keV provides a good fit to the data.  
 However, the strong FeK line at 6.7 keV which should accompany the 
 thermal emission is not observed. This could be attributed to a low
metallicity, but unfortunately we cannot check this as the abundances
are not well constrained using the present X-ray data.
  Alternatively, the lack of a FeK line could be explained by a low
 contribution of type Ia supernovae to the enrichment of the 
Interstellar Medium (Sansom et al. 1996). 

  Finally we consider high mass 
 X-ray binaries as a possible origin of the hard X-ray component. 
  High mass binaries will form as a consequence of the 
 starburst, and indeed many such systems have been 
 identified in nearby star-forming galaxies (Read et al. 1997, 
  Fabbiano 1995 and references therein).  
 Assuming that a typical X-ray luminosity of these systems is
 $10^{37-38}$ $\rm erg~s^{-1}$, we can estimate the total number of 
 binaries. As the X-ray luminosity of the hard component is 
 $L_x\sim 5\times 10^{41}$ and $L_x\sim 10^{41}$ $\rm erg~s^{-1}$   
 for NGC3690 and NGC3310 respectively, we estimate a range of between 
 5,000-50,000 and 1,000-10,000 X-ray binaries for the two 
 galaxies. Now we can compare this with the number of ionizing OB stars
 determined from the integrated far-infrared luminosity. Making the 
 assumption that it is mostly these stars that heat the dust, which then
 reradiates in the mid-far infrared, we estimate $\sim\ 2\times 10^{5}$
 and $\sim\ 3.5\times 10^{6}$ OB stars for NGC3310 and NGC3690,
 respectively. If, following Fabbiano et al. (1992), 0.2 percent of these
 are massive X-ray binaries, then there are 400, and 7000 such systems,
 respectively. For the upper range of binary luminosities the
 predicted X-ray luminosity is comparable
 to that observed in both cases. Indeed, if the hard  
 X-ray emission arises from binaries with 
 low-metallicity and thus high X-ray luminosity  
 ($L_x\sim10^{38-39}$ $\rm erg~s^{-1}$) like those 
 observed in the Magellanic Clouds (van Paradijs \& McClintock 1995),
 then they could easily produce the observed luminosity. 
  One potential problem is that some point sources 
  observed in nearby galaxies with the ROSAT PSPC by Read et al. (1997),
  appear to have soft spectra ($\rm kT\sim 2$ keV), whilst the
 the high temperatures inferred for NGC3310 and NGC3690 are closer 
 to those predicted for high mass X-ray binaries (see Nagase 1989).

\section{CONCLUSIONS}

We have modelled and interpreted the combined ASCA and ROSAT X-ray
spectra from 0,1-10 keV, for the star-forming galaxies NGC3310 and
NGC3690. 
These two galaxies are amongst the most X-ray luminous 
 ($L_x\sim 10^{41-42}$) $\rm erg~s^{-1}$ starbursts in the local Universe. 
 In addition, ROSAT HRI images show that
the emission from NGC3310 is extended out to at least 
 $\sim$ 3 kpc. The soft X-ray emission from NGC3690 comes from at least 
 three spatially resolved regions. The limited spatial resolution of 
 the ASCA data does not allow us to 
 place useful limits on the extent of their hard X-ray components.  
 The X-ray spectrum of NGC3310 can be described by two components:  
 at soft energies a Raymond-Smith component (kT$\sim$0.8 keV) 
 which  probably 
 originates from a super-wind. The predicted soft X-ray emission,  
 on the basis of the supernovae mechanical energy deposited to 
 the interstellar medium, 
 is comparable to that observed.
 At harder energies we can fit either a Raymond-Smith 
 component (kT$\sim$17 keV) or a power-law $\Gamma\sim 1.4$. 

 The results for NGC3690 are similar. There are at least two 
 components in its spectrum: a soft Raymond-Smith plasma 
  (kT$\sim$0.8 keV) and a harder component which can be 
 represented equally well 
 by a Raymond-Smith component (kT$\sim$10 keV) or a hard 
 power-law $\Gamma \sim 1.6$. 
  However, the best-fit model for NGC3690 is rejected at the 98 per cent 
 confidence level implying that more complicated models are necessary.
 Hence we considered a non-equilibrium 
 ionization model (Hughes et al. 1994) 
  and a multi-temperature thermal model (Done \& Osborne 1996). 
 The above models do improve the fit, but only at the $\sim90$ per 
 cent confidence level. Although non-equilibrium models may be 
 important, additional information such as X-ray emission line ratios
 are required in order to make further quantitative progress.

  In neither NGC3310 and NGC3690 do we find evidence for 
 significant absorption columns above the Galactic values,
 in either the soft or the hard X-ray component.

 The nature of the 
 hard X-ray emission is still uncertain, and we considered various
 possibilities.
 Of these we conclude that Inverse Compton scattering by 
 high energy radio electrons and infrared photons, can probably 
 provide only a minor contribution. Although the presence of an AGN
 is a possibility, this is not supported by data at
 other wavelengths, and the modest columns argue against an obscured
 nucleus. A very hot thermal component may be present, ($\sim10^{8}$ K).
 We do not detect the FeK line at 6.7 keV, possibly because the 
 the abundances are sub-solar.
 To test this
 possibility further requires information on element abundances, and 
 an extension of the observed spectra to higher energies.  Finally, 
 estimates of the number of X-ray binaries, based on the total mid-IR emission,
 suggests that they may well be able to account for the hard X-ray 
 emission. 
 
 The X-ray results on these two galaxies are similar to those
 obtained for other dwarf star-forming galaxies 
 (eg NGC1569, NGC4449) and  the archetypal star-forming 
 galaxies M82 and NGC253, although there is a range in relative contributions
 from the various X-ray components.
 This implys that the same general emission mechanisms 
 apply in star-forming galaxies over three decades of luminosity. 
 
\section{Acknowledgments}

We are gratefull to J. Hughes for providing the supernova remnant 
 non-equilibrium XSPEC model.

\section*{References}
Allen D.J.,  ASTERIX User Note 004, STARLINK, 1992 \\
Armus L., Heckman T.M., Miley G.K., 1990, ApJ, 364, 471 \\
Bade N., Komossa S. and Dahlem M., 1996, A\&A, 309, 35L \\
Balick B. and Heckman T., 1981, A\&A, 96, 271\\
Bertolla A.F., and Sharp N.A., 1984, MNRAS, 207, 47 \\
Bevington P.R. and Robinson D.K., 1992, Data reduction and error analysis for the
physical sciences, 2nd ed., McGraw Hill \\
Bohlin R.C., Savage B.D. and Drake J.F., 1978, ApJ., 224, 132\\ 
Casoli F., Combes F., Augarde R., Figon P. and Martin J.M., 1989,
A\&A, 224, 31 \\
David \etal 1997 The \rosat HRI Calibration Report \\ 
hea-www.harvard.edu/rosat/rsdc\_www/HRI\_CAL\_REPORT
\\
de Vacouleurs G. \etal, 1991, Third Reference catalogue of bright
galaxies. \\
Dekel A. and Silk J., 1986, ApJ, 303, 39 \\
Della Ceca R., Griffiths R.E., Heckman T.M. and Mackenty J.W., 1996, ApJ, 469, 662 \\
Della Ceca R., Griffiths R.E. and Heckman T.M., 1997, ApJ, 485, 581
\\
Done C.D. and Osborne J.P., 1997, MNRAS, 288, 649 \\
Elvis M., Soltan A. and Keel W., 1984, ApJ., 283, 479
Fabbiano G., 1988, ApJ, 3310, 672 \\
Fabbiano G., 1995, in X-ray Binaries edited by
Lewin W.H., van Paradjis J., and van den Heuvel E.P.J. \\ 
Fabbiano G. and Trinchieri G., 1984, ApJ, 286, 491 \\
Fabbiano G., Kim D.W., and Trinchieri G., 1992, ApJS, 80, 531 \\
Fabbiano G., Schweizer F. and Mackie G., 1997, 478, 542 \\
Friedman S.D., Cohen R.D., Jones B., Smith H.E. and Stein W.A., 1987,
AJ, 94, 1480 \\
Gehrz R.D., Sramek R.A., Weedman D.W., 1983, ApJ, 267, 551 \\
Gendreau K.C., 1995, PhD. Thesis \\
Griffiths R.E. and Padovani P., 1990, ApJ., 360, 483 \\
Heckman T.M, Armus L.,Miley G.K., 1990, ApJS, 74, 833 \\
Heckman T.M, Dahlem M, Lehnert M.D., Fabbiano G., Gilmore D., Waller
W.H., 1995, ApJ., 448, 98\\
Ho L.C., Filippenko A.V., Sargent W.L., 1995, ApJS, 98, 477 \\
Ho L.C., Filippenko A.V., Sargent W.L.W., 1997, APJS, 112, 315 \\
Hughes J.P. and Singh K.P., 1994, ApJ, 422, 126 \\
Kim D.W., Fabbiano G., Trinchieri G., 1992, ApJS, 80, 645 \\
Lehnert M.D. and Heckman T.M., 1996, ApJ, 462, 651 \\
Long K.S. and van Spreybroeck L.P., 1983, in Accretion driven stellar
X-ray sources, edited by Lewin W.H.G. and van Den Heuvel E.P.J., CUP
\\
Longair M.S. 1992, High Energy Astrophysics vol.2,  CUP  \\
MacLow M.M., and McCray R., 1988, ApJ, 324, 776 \\
Moran E.C., Halpern J.P., Helfand D.J., 1996, ApJS, 106, 341 \\ 
Moran E.C. and Lehnert M.D., 1997, ApJ, 478, 172 \\
Mulder P.S. and Van Driel W., 1996, A\&A, 309, 403 \\
Nagase F., 1989, PASJ, 41, 1 \\
Nakagawa T., Nagata T., Geballe T.R., Okuda H., Shibai H., Matsuhara
H., 1989, ApJ, 340, 729 \\
Niklas S., Klein U., Weilebinski R., 1997, A\&A, 322, 19 \\
Ohashi \etal, 1990, ApJ, 365, 180 \\
Ohashi \etal, 1996, PASJ, 48, 157 \\
Pastoriza M.G., Dottori H.A., Terlevich E., Terlevich R., Diaz A.I.,
1993,MNRAS, 260, 177 \\
Pfefferman \etal, 1986, Proc. SPIE, 733, 519\\
Ptak A., Serlemitsos P., Yaqoob T., Mushotzky R., Tsuru T., 1997, AJ,
113, 1286 \\
Raymont J.C., Smith B.W., 1977, ApJSS, 35, 419 \\
Read A.M., Ponman T.J., Strickland D.K., 1997, MNRAS, 286, 626 \\
Read A.M., Ponman T.J., Wolstencroft R.D., 1995, MNRAS, 277, 397 \\
Rieke G.H., Lebofsky M.J., Thompson R.I., Low F.J., Tokunaga A.T.,
1980, ApJ., 238, 24 \\
Roberts T., 1998, PhD Thesis, University of Leicester, in preparation \\
Sanders D.B. and Mirabel I.F., 1985, ApJ., 298, 31L \\
Sansom A.E., Dotani T., Okada K., Yamashita A., Fabbiano G., 1996,
MNRAS, 281, 48 \\
Schaaf R. \etal, 1989, ApJ, 336, 722\\
Schultz H. \etal, 1997, preprint\\
Serlemitsos P., Ptak A., Yaqoob T., 1996, in The physics of LINERS in
view of recent observations, 1996, edited by M. Eracleous,
A. Koratkar, C. Leitherer, and L. Ho \\
Smith \etal ,1996, ApJ, 473, L21 \\
Soifer B.T. \etal , 1987, ApJ, 320, 238 \\
Sutherland R.S., Dopita M.A., 1993, ApJS, 88, 253 \\
Stark A. A. \etal, 1992, ApJS, 79, 77\\
Stewart G.C., Fabian A.C., Terlevich R.J., Hazard C., 1982, MNRAS,
200, 61 \\ 
Strickland D.K., Ponman T.J., Stevens I.R., 1997, A\&A, 320, 378 \\
Tanaka Y., Inoue H., Holt S.S., 1994, PASJ, 46, 37L \\
Telesco C.M. and Gatley I., 1984, ApJ, 284, 557 \\
Tr\"{u}mper J., 1984, Physica Scripta, T7, 209  \\
Tsuru T.G., Awaki H., Koyama K., Ptak A., 1998, PASJ, 49, 619 \\
Vall\'{e}e J.P., 1993, MNRAS, 264, 665 \\
Van der Kruit P.C. and de Bruyn A.G., 1976, 48, 373 \\
Van Paradjis J. and McClintock J.E., 1995, in X-ray Binaries edited by
Lewin W.H., van Paradjis J., and van den Heuvel E.P.J. \\
Veilleux S., Osterbrock D.E., 1987, ApJS, 63, 295 \\
Voges W. \etal, 1996, IAU Circ. 6420 \\
Watson M.G., Stanger V., Griffiths R.E., 1984, ApJ, 286, 144  \\
Wilson A.S.\etal, 1992, ApJ, 391, L75 \\
Wynn-Williams \etal., 1991, ApJ., 377, 426 \\
Yaqoob T. \etal, 1995, ApJ, 455, 508 \\
Yaqoob T. \etal, 1997, The \asca ABC Guide, v2.0, NASA/GSFC  \\
Zhao J.H., Anantharamaiah K.R., Goss W.M., Viallefond F., 1997, ApJ, 482,
186 \\

\end{document}